\documentclass[preprint,aps,showpacs]{revtex4}

\usepackage{graphicx}

\usepackage{bm}

\begin{document}

\title{Incoherent scattering of highly ultrarelativistic

channeled particles on electrons}

\author{Victor V. Tikhomirov}

\affiliation{Research Institute for Nuclear Problems, Belarusian
State University, Minsk, Belarus}

\date{\today}

\begin{abstract}
This article systematically examines the effects that must be
taken into account when modeling the dechanneling of highly
ultrarelativistic charged particles caused by incoherent
scattering by electrons of crystal atoms. Changes in the particle
electromagnetic field that occur in condensed matter at highly
ultrarelativistic particle energies are taken into account. The
excitation of both collectivized and inner-shell electrons of
atoms is considered, taking into account thermal vibrations and
the inhomogeneous periodic distribution of atoms in crystals.
Differences in the calculations of stopping power and mean square
scattering angle by electron of channeled particles are described.
The need to separate both of them into diffuse processes,
accompanied by the transmission of a transverse momentum which
does not exceed the transverse momentum of channeled particles,
and instantaneous dechanneling processes with a greater
transmitted momentum, the variation range of which reaches several
orders of magnitude, is justified. The main result of the present
work is expressions for the mean square projected scattering
angles per unit length in planes parallel and perpendicular to the
atomic plane, which also allow for the introduction of effective
minimum momenta for incoherent scattering. Expressions of both
types take into consideration completely the nonlocal effects of
momentum transfer from crystal electrons to particles moving along
classical trajectories, and allow for the formulation of simple
methods for modeling the ultrarelativistic particle incoherent
scattering by electrons.
\end{abstract}


\maketitle

\section{Introduction}
The phenomenon of charged particle channeling along the directions
of crystal planes and axes, which is also possible in other
ordered structures, has been studied since the 1960s
\cite{bar,bir}. Applications of channeling include the study of
associated  radiation \cite{bar} and narrow-band radiation in
crystalline undulators \cite{bar2, ksg, bel}, as well as the
measurement of the electromagnetic characteristics of particles
based on the spin precession effect \cite{bar3, kim, fom, aio}. In
addition, stable channeling in thick crystals is important for the
development of methods for accelerating particles to ultrahigh
energies \cite{zim} and ensuring the extraction of particle beams
from storage rings \cite{els}. The efficiency of these
applications increases with increasing crystal thickness, limited
only by the dechanneling process, the main mechanism of which is
the incoherent scattering of particles on the electrons of the
crystal atoms, hereinafter referred to as $\it
{electron~scattering}$.

The specifics of incoherent electron scattering under channeling
conditions have been considered primarily in relation to the
calculation of stopping power \cite{oht, nit, esb, bur, esb2},
while the use of channeling in high-energy physics also requires a
theory of electron dechanneling of ultrarelativistic particles,
which has not yet been formulated independently. In particular,
the entire cycle of studies on the use of crystals for collimating
storage ring beams \cite{st} was accompanied by an empirical
description of incoherent electron scattering based on the
well-known formula for stopping power in a homogeneous amorphous
medium \cite{pdg}.

However, the application of this formula to the description of
channeling, as well as the formulas of the theory \cite{oht, nit,
esb, bur, esb2}, require significant modification. Firstly, in
these studies, when calculating the stopping power of channeled
particles, integration was performed over all energy transmissions
to electrons up to the maximum one, which at sufficiently high
energies can be comparable to the total energy of the particle.
However, starting from an energy transmissions several orders of
magnitude lower, electron scattering of particles of sufficiently
high energies causes instantaneous dechanneling. Therefore, when
calculating continuous energy losses and the mean square angle of
diffuse scattering of particles in the channeling regime, only
limited energy transmissions should be taken into account,
treating the larger ones as single scattering, leading, as a rule,
to instantaneous dechanneling. The unjustified inclusion of all
possible energy transmissions in the calculation of the stopping
power and mean scattering angle of channeled particles leads to an
overestimation of these values by 50 \% or more, to compensate for
which a phenomenological coefficient was used in the works
\cite{bir, st}. Indeed, the transfer of energy $T$ and transverse
momentum $k_\perp$ to a free electron at rest are connected by the
relativistic relation  $T = k_\perp^2/2m$ over a wide range of
transmitted energies, where $m$ is the electron mass. This
relation leads to a proportionality coefficient of $2m/p^2$
between the stopping power and the mean square scattering angle.
However, in \cite{st} and \cite{bir} instead of it, the factors
$m/p^2$ and $m/2p^2$ were used, respectively, which was the only
thing that could compensate for the overestimation of the stopping
power of channeled particles.

Furthermore, in \cite{bir, st}, the formula for the stopping power
is used in conjunction with the local electron density
approximation, which is valid only for momentum transfers of the
order of $\hbar/2\cdot 10^{-10} cm \simeq 100 keV/c$ and higher.
It is known from the theory of stopping power \cite{mol, bet, ll3,
ll4, ll8, tm} that smaller momentum transmissions, described using
the average electron density of the medium and taking its
dispersion into account, are also important in electron
scattering. It is clear that there is also an intermediate region
of momentum transmissions at impact distances from several
hundredths to several interatomic spacings, which, depending on
the transverse coordinates of the particle, includes regions with
different electron density and binding energy distributions. In
this case, the description of scattering in crystals requires
taking into account the inhomogeneous periodic distributions of
both the electron density and binding energy. Although the idea of
solving this problem was proposed back in \cite{esb}, devoted to
calculating the stopping power of nonrelativistic channeled
particles, its application to the case of electron scattering of
ultrarelativistic particles requires additional consideration of a
number of factors. As already noted, it is necessary to consider
the possibility of particle dechanneling over a wide range of
energy transmissions to electrons, as well as the influence of
thermal vibrations of atoms, which lead to the suppression of
large coherent momentum transfers to the periodic atomic lattice.
Furthermore, when calculating the mean square multiple scattering
angle, it is necessary to take into account that it is associated
with the transmission of the transverse component of momentum,
while the energy transmission to electrons, both at its minimum
and maximum values, is mostly associated with the transmission of
longitudinal momentum.

In this paper, we calculate the mean square angle of incoherent
scattering by electrons along the trajectory of an
ultrarelativistic particle, taking into account the periodicity
and nonuniformity of the arrangement and thermal vibrations of
crystal atoms, the modification of the electromagnetic field of
ultrarelativistic particles in a dense medium, the collective
nature of the excitations of valence and conduction band
electrons, and instantaneous dechanneling accompanying large
energy transmissions. The resulting expressions describe the
nonlocal nature of scattering and can be used both to describe the
diffuse scattering of channeled particles and to introduce an
effective minimum momentum transmission, which can be used in
Monte Carlo algorithms for modeling the applications of
high-energy particle channeling effect.

\section{Semiclassical expression for the mean square angle of incoherent multiple electron scattering}
The theory of incoherent electron scattering of high-energy
particles must combine the classical description of their motion
with a quantum calculation of the incoherent transmission of
momentum to the crystal atoms. Let us consider the excitation of a
medium by a relativistic particle with a velocity $\bm v$, the
scalar $\varphi$ and vector $\bm A$ potentials of which are
determined by the Lorentz equations \cite{ll8, tm} (the article
uses the system of units $c = \hbar = 1$)
\begin {equation} 
\begin{array}{c}
\displaystyle{ \hat{\varepsilon} \bigg(\Delta \varphi -
\hat{\varepsilon} \frac{\partial^2 \varphi}{\partial t^2}\bigg) =
-4 \pi e \delta
(\bm R - \bm R_p) }, \\
\\
\displaystyle{ \Delta \bm A - \hat{\varepsilon} \frac{\partial^2
\bm A}{\partial t^2} = -4 \pi e \bm v \delta (\bm R - \bm R_p) },
\end{array}
\end{equation}
where $\bm R_p = \bm b + \bm v t$  is the radius-vector of a
particle colliding with an atom with an impact parameter $\bm b
\perp \bm v$, $\hat{\varepsilon}$ is the permittivity operator
\cite{ll8}, which takes into account both the spatial and temporal
dispersion described by the dependence of its Fourier component
$\hat{\varepsilon}(\bm k, \omega)$   on the momentum (wave vector)
and frequency.

The solution to the system of equations (1) has the form
\cite{ll8}
\begin {equation} 
\begin{array}{c}
\displaystyle{ \varphi(\bm R, \bm b, t) =4 \pi e \int \frac{1}{\varepsilon(\bm k, \omega)} \frac{1}{k^2-\varepsilon(\bm k,
\omega)\omega^2} \: e^{i \bm k (\bm R - \bm b - \bm v t)} \frac{d^3 k}{(2 \pi)^3} }, \\
\displaystyle{ \bm A(\bm R, \bm b, t) = 4 \pi e \int \frac{\bm
v}{k^2-\varepsilon(\bm k, \omega)\omega^2} \: e^{i \bm k (\bm R -
\bm b - \bm v t)} \frac{d^3 k}{(2 \pi)^3} },
\end{array}
\end{equation}
where $\bm R = \bm r + \bm r_a$ is the radius vector of the
electron, equal to the sum of the radius vector $\bm r$ of its
position relative to the nucleus of the atom and the radius vector
$\bm r_a$ of the latter, which has a projection $\bm \rho_a \perp
\bm v$ onto the impact parameter plane. The Hamiltonian of the
interaction of a particle with an atomic electron has the form
\cite{ll3, ryab}
\begin {equation} 
\hat{H}(\bm R,\bm b,t)= \hat{H}_{at}(\bm R) + e\cdot \varphi (\bm
R,\bm b,t) - \bm j \cdot \bm A (\bm R,\bm b,t)
\end{equation}
where $\bm j = i(\bm \nabla e^{i\bm k \bm r} - e^{i\bm k \bm r}
\bm \nabla )/2m$ is the probability density flux,
$\hat{H}_{at}(\bm R)$  is the Hamiltonian of an electron in an
atom with eigenstates $|n\rangle = \psi_n(\bm r)$, $n = 1, 2,..$
and eigenvalues $E_n$. The action of the particle field (2)
excites the electron from the ground state $\psi_0(\bm R)$ to
\cite{ll3, ryab}
\begin {equation} 
\Psi(\bm R,\bm b,t)= \psi_0(\bm R) e^{-iE_0 t} + \sum_{n\neq 0}
a_{n0} (\bm \rho_a,\bm b)\psi_n(\bm R) e^{-iE_n t},
\end{equation}
where
\begin {equation} 
\begin{array}{c}
\displaystyle{ a_{n0}(\bm \rho_a,\bm b) = -i \int_{-\infty}^\infty
dt (e\varphi -
\bm A \bm j )_{n0} e^{i \omega_{n0}t} } \\
\displaystyle{= -i \int_0^{k<k_1} \frac{d^2 k}{4 \pi^2}
\int_{-\infty}^\infty d\omega M_{n0} (\bm k, \omega) e^{\bm k (\bm
\rho_a - \bm b)} \delta(\omega - \omega_{n0}) },
\end{array}
\end{equation}

\begin {equation} 
M_{n0}(\bm k, \omega) = 4 \pi e^2 \frac{\langle n |e^{i \bm k \bm
r}-i \bm v \varepsilon(\bm k, \omega)(\bm \nabla e^{i\bm k \bm r}
- e^{i\bm k \bm r} \bm \nabla )/2m |0\rangle}{\varepsilon(\bm k,
\omega)(k^2 - \varepsilon(\bm k, \omega)\omega^2 )},\hspace{1 cm}
\omega = \bm k \bm v.
\end{equation}
Note that in (3), in accordance with \cite{bet, ll3, ll4}, an
upper limit $k_1$ of integration over the modulus of the momentum
$\bm k$ is introduced.

Using the wave function (4) and the transverse momentum operator
$\hat{\bm p} = -i \partial / \partial \bm b$ of the particle, we
introduce the averaged product of the transverse components
\begin {equation} 
\big\langle \hat{p}_i \check{p}_k \big\rangle = \int \hat{p}_i^*
\sum_a \Psi^* (\bm r, \bm \rho_a, \bm b, t)\: \check{p}_k \sum_b
\Psi (\bm r, \bm \rho_b, \bm b, t) d^3 r
\end{equation}
 of the
momentum \cite{ryab} transmitted to the particle during the
interaction with the set of atoms over which the summation is
explicitly designated, while the summation sign for the electrons
of each atom is traditionally omitted \cite{ll4}. Also, function
(4) contains the summation over the excited states of the atomic
electron.

Let's start with a situation of randomly uniformly distributed
atoms, which serves as a model for both an amorphous medium and a
misoriented crystal. Since scattering in this situation is
azimuthally symmetric with respect to the particle velocity, it is
natural to consider the mean square of the nonprojected (space)
\cite{pdg} scattering angle, summed over the transverse
directions. A uniform distribution over the impact parameter $\bm
b$ allows integration over it (see (11) below) to obtain
\begin {equation} 
\begin{array}{c}
\displaystyle{\bigg\langle \frac{d
p^2_\perp(k_1)}{dz}\bigg\rangle_{\bm g = 0} = \frac{4e^4 N Z}{v^2}}\\
\\
\displaystyle{\times \int_0^{k<k_1} k^2_\perp d^2 k_\perp
\int_0^\infty d \omega \sum_{n\neq 0}\frac{|\langle n|e^{i\bm k
\bm v}-i \bm v \varepsilon(\bm k, \omega) ( \bm\nabla e^{i\bm k
\bm v} - e^{i\bm k \bm v} \bm\nabla )/2m| 0 \rangle|^2}{|k^2_\perp
+\omega^2/v^2-\varepsilon(\bm k,
\omega)\omega^2|^2}\frac{\delta(\omega-\omega_{n0})}{|\varepsilon(\bm
k, \omega)|^2}},
\end{array}
\end{equation}
where $N$ is the number density of the substance's atoms and $NZ$
is the same of electrons.

Moving on to the case of a crystal, we represent the sum over its
atoms as a product \cite{tm, bks}
\begin {equation} 
\sum_a e^{i \bm k, \bm \rho_a} = S_{cr}(\bm k) \sum_{i_{lat}} e^{i
\bm k, \bm \rho_{i_{lat}} },
\end{equation}
of the sum over the nodes of the crystal lattice and the
structural factor
\begin {equation} 
S_{cr}(\bm k) = \sum_{j=1}^{s_{cell}} e^{i \bm k, \bm \rho_j},
\end{equation}
in which the summation is carried out over the atoms of the unit
cell. The index "cr" is introduced here to distinguish the
structure factor $S_{cr}$ from the scattering function, also
traditionally denoted by the symbol $S$ \cite{hub}. We will also
carry out averaging over the thermal vibrations of the atoms,
adding for this the corresponding deviations from the equilibrium
positions and, in accordance with the simplest model of thermal
vibrations, averaging over the Gaussian distribution of these
deviations with the root-mean-square amplitude $u_1$ \cite{tm,
bks}.

When summing over the crystal atoms, it must be borne in mind that
the coordinate dependence of the incoherent scattering intensity
is significantly manifested only under channeling conditions and
in its close vicinity, while for motion at large angles to the
crystal axis or plane, the characteristics averaged in the
transverse plane (impact parameter plane) are quite sufficient.
Taking into account longitudinal momentum transmissions is also
necessary to ensure the transition to the description of particle
scattering on individual atoms \cite{maz} at angles much greater
than the channeling angle, however, at the angles of the order of
the latter, such transmissions are exponentially suppressed
\cite{maz}, and we will neglect them. As a result, the sum over
the longitudinal coordinates of the lattice sites will give the
ratio of the length $\Delta z$  of a rectilinear section of the
particle trajectory, parallel to the chosen crystallographic
direction, to the interatomic spacing along it. As a result of
averaging over the thermal vibrations of the atoms and summing
over the crystal lattice sites using the method \cite{tm}, we
obtain
\begin {equation} 
\bigg\langle \sum_a e^{i(\bm k_1 - \bm k_2) \bm \rho_a}
\bigg\rangle = \frac{4 \pi^2 \Delta z}{\Omega} \sum_{\bm g}
S_{cr}(\bm g) e^{-g^2 u_1^2/2} \delta^2 (\bm k_{1\bot} - \bm
k_{2\bot} -\bm g),
\end{equation}
where $\Omega$ is the volume of the unit cell, equal $a^3$ in the
case of a cubic cell with an edge length $a$ and related to the
number of crystal atoms per unit volume $N$ by the relation
$S_{cr}(0) = N \Omega$.

After substituting (11) and (4)-(6), expression (7) for the
average product of the transverse components of the momentum
\cite{ryab}, associated with incoherent electron scattering over a
unit length of the particle trajectory with an impact parameter
$\bm b$, takes the form
\begin {equation} 
\bigg\langle \frac{d p_i p_j (\bm b)}{dz}\bigg\rangle = \frac{1}{4
\pi^2 v^2 \Omega} \sum_{\bm g} S_{cr}(\bm g) e^{-g^2 u_1^2/2}e^{i
\bm g \: \bm b} \int d^2 k_\perp (k_{+i}k_{-j}) \sum_{n=1,..}
M^*_{n0}(\bm k_-) M_{n0}(\bm k_+)~,
\end{equation}
where the summation over the electrons of one atom is still
implied and vector combinations
\begin {equation} 
\bm k_\pm = \bm k_\bot \pm \frac{1}{2}\:\bm g
\end{equation}
are introduced. It is easy to verify that the term with $\bm g$ =
0 corresponds to the case of an amorphous medium (8).

\section{Integration over the transmitted momentum in the case of an
amorphous medium}

Let us begin by calculating the mean square scattering angle in a
medium of randomly uniformly distributed atoms, which serves as a
model of both an amorphous medium and a disoriented crystal,
rewriting equation (8) it in a more detail
\begin {equation} 
\begin{array}{l}
\displaystyle{\bigg\langle \frac{d
p^2_\perp(k_1)}{dz}\bigg\rangle_{\bm g = 0} = \frac{4e^4}{v^2}
\int_0^{k<k_1} } k^2_\perp d^2 k_\perp \int_0^\infty d \omega
\sum_{n\neq 0, NZ}\frac{\delta(\omega-\omega_{n0})}{|k^2_\perp
+\omega^2/v^2-\varepsilon(\bm k, \omega)\omega^2|^2}\\
\times \displaystyle{\left\{ \frac{|\langle n|e^{i\bm k \bm v}| 0
\rangle|^2}{|\varepsilon(\bm k, \omega)|^2}  + 2 \Re \bigg(
\frac{\langle n|e^{i\bm k \bm v}| 0 \rangle^* \langle n|-i \bm v (
\bm\nabla e^{i\bm k \bm v} - e^{i\bm k \bm v} \bm\nabla | 0
\rangle}{2m \varepsilon(\bm k, \omega)}\bigg) \right.
}\\
\displaystyle{\left. + \frac{|\langle n|\bm v ( \bm\nabla
e^{i\bm k \bm v} - e^{i\bm k \bm v} \bm\nabla
| 0 \rangle |^2}{4 m^2} \right\} }\\
\displaystyle{\simeq \frac{4e^4}{v^2} \int_0^{k<k_1} k^2_\perp d^2
k_\perp \int_0^\infty d \omega \sum_{n\neq 0, NZ}\frac{|\langle
n|e^{i\bm k \bm v}| 0 \rangle|^2}{|k^2_\perp
+\omega^2/v^2-\varepsilon(\bm k, \omega)\omega^2|^2 }
\frac{\delta(\omega-\omega_{n0})}{|\varepsilon(\bm k, \omega)|^2},
}
\end{array}
\end{equation}
where, as before, a non-relativistic boundary integration impulse
$k_1$ is used, significantly exceeding the threshold impulse of
electronic excitation \cite{bet, ll3, ll4}.

Let us consider the processes of momentum transmission to
collective excitations of the medium and electrons of the internal
atomic shells using the corresponding representations of the
dynamic structure factor \cite{pin, ege}
\begin {equation} 
S(\bm k, \omega) =  \sum_{n\neq 0} |\langle n|e^{i\bm k \bm v}| 0
\rangle|^2 \frac{\delta(\omega-\omega_{n0})}{|\varepsilon(\bm k,
\omega)|^2} = \frac{k^2}{4 \pi^2 e^2} \Im
\frac{-1}{\varepsilon(\bm k, \omega)}
\end{equation}

\subsection{Excitation of electrons of the valence band of
semiconductors}

The excitation of both conduction band electrons in conductors and
valence band electrons in semiconductors is qualitatively
described by the Fermi gas model. The Fermi energies, plasma
frequencies and energy gap widths \cite{br} of the monoatomic
semiconductor crystals are given in the table below

\vspace{5mm} \hspace{4cm}
\begin{tabular}{|c|c|c|c|}
\hline
& ~~$\varepsilon_F$, eV~~ & ~~$\omega_p$, eV ~~& ~~$E_g$, eV~~ \\
\hline
~C~ & 25 & 28 & 13.6 \\
\hline
~Si~ & 13 & 17 & 4.8 \\
\hline
~Ge~ & 12 & 16 & 4.2 \\
\hline
\end{tabular}
\vspace{3mm}

\hspace{2cm}Fermi energies, plasma frequencies and energy gap
widths from \cite{br}. \vspace{3mm}

Note that the parameter of energy gap $E_g$ from \cite{br} differs
from the energy distance separating the valence band and the
conduction band in semiconductors, and is determined in \cite{br}
by the static value of the substance permittivity .

Plasmons \cite{pin, ege, br, bp, lin, fer} play a leading role in
the excitation of the Fermi gas at low energies and momentum
transfers. They are most easily excited when $\omega$ exceeds the
plasma frequency $\omega_{pv}$, calculated for the valence band
electron number density $n_v$, and the transverse momentum
transfer does not exceed a certain threshold value $k_c \sim 0.7
k_F$, where $k_F$ is the Fermi momentum \cite{pin, ege, br, bp,
lin, fer, glk, rae}.

Single-particle electron-hole excitations of the Fermi gas are
possible starting from zero energy in conductors and the energy of
the gap width in semiconductors. However, for momentum
transmission $k_\perp < k_c$, they are strongly suppressed by the
screening effect described by the square of the permittivity
modulus in the denominator of (15) \cite{pin}. In addition, the
significant value of the imaginary part of the permittivity
\cite{phil} also leads to the suppression of energy losses in
semiconductors in the region of frequencies somewhat lower than
the plasma frequency, at which the condition for the occurrence of
Cherenkov radiation is satisfied for ultrarelativistic particles.
However, at $k_\perp > k_c$, the probability of plasmon decay into
single-particle excitations increases sharply, which begin to play
a major role, quickly becoming similar to the scattering on free
electrons \cite{pin, ege, fer, glk, rae}.

The described picture of excitation of the Fermi gas of electrons
in the valence band of semiconductors and the conduction band of
conductors has been developing since the 1950-s to the present day
\cite{pin, ege, br, bp, lin, fer, glk, rae, phil,lun, rh, penn,
penn2, shin, da}. Recent model \cite{shin, da} uses
$\it{full~Penn~algorithm}$, based on the phenomenological
implementation of Lindhardt's theory \cite{lin} in the
permittivity formalism \cite{pin} and require specific
calculations involving experimental data.

In the case of high particle energies, the peculiarities of
scattering on valence electrons manifests itself in a relatively
small region of transmitted momenta and, in the logarithmic
approximation, can be taken into account by the simple single-mode
excitation model \cite{br, penn, penn2}, which describes the
excitation processes of plasmons and electron-hole pairs by a
single formula. Despite its simplicity, this model agrees well
with more complex theories and experimental data \cite{penn,
penn2}. In this approach, the formula
\begin {equation} 
\Im \frac{-1}{\varepsilon(\bm k, \omega)} = \frac{\pi}{2}
\frac{\omega_{pv}^2}{\omega_{\bm k}} \delta(\omega-\omega_{\bm
k}),
\end{equation}
introduced to describe the excitation of plasmons, is supplemented
by a generalized expression for the excitation energy
\begin {equation} 
\omega_{\bm k}^2 = \omega_{pv}^2 +E^2_g + \alpha \frac{k^2_F}{m^2}
k^2 + \frac{1}{4m^2} k^4~.
\end{equation}
Unfortunately, the parameter $\alpha$ before the penultimate term
is not very clearly defined. Specifically, the value $\alpha =
3/5$ is used in \cite{ege, br, fer, rh}, while the value $\alpha =
1/3$ - in \cite{lun, penn, penn2, shin} and in this paper.

Substitution of (16), (17) into (14) leads in the limit $k_F \ll
k_1$, $\omega_{pv} \ll m$ both to a logarithmic dependence
\begin {equation} 
\begin{array}{l}
\displaystyle{\bigg\langle \frac{d
p^2_\perp(k_1)}{dz}\bigg\rangle_v = \frac{4\pi e^4 n_v}{v^2}\:
ln\frac{k_1^2}{\alpha k^2_F + m \sqrt{\omega_{pv}^2 + E^2_g} }
}\\
\displaystyle{= \frac{4\pi e^4 n_v}{v^2} \: ln\frac{k_1^2}{2 m
I_v} },\\
\displaystyle{I_v = \alpha E_F + \frac{1}{2}\sqrt{\omega_{pv}^2 +
E^2_g} }
\end{array}
\end{equation}
on the upper limit $k_1$ of the transmitted momentum and to the
introduction of effective ionization momentum $\sqrt{2 m I_v}$,
the value of which is simultaneously determined by the width of
energy gap (if any), plasma frequency and Fermi energy of valence
or conduction band electrons.

It should be clarified that taking into account the current
contribution in Hamiltonian (3) leads to the appearance in (14) of
the terms that are absent in nonrelativistic theory \cite{ll3},
which we can neglect when describing the contribution of
collective excitations in the logarithmic approximation. Indeed,
due to the significant value of the square of the permittivity
modulus at excitation energies of several electron volts
\cite{phil}, the effective lower limit of integration of the last,
squared in the current contribution, term of (14) turns out to be
large enough to use the plane wave approximation for estimating
the matrix elements in the logarithmic approximation. This allows
us to verify that the contribution of this term contains a factor
$k_1^2/m^2$, which, by virtue of the condition $k_1 \ll mc$, has
the order of negligible relativistic correction. The second term
in the curly bracket (14) describes the interference of the
"plasmonic" and current contributions. In addition to the
smallness of the latter, the possibility of neglecting the
integral of this term is additionally facilitated by the fact that
it is proportional to the real part of the permittivity, which
passes through zero at the plasma frequency.

\subsection{Excitation of electrons of inner atomic shells}

Plasma excitations play a leading role in both energy losses and
incoherent scattering of electrons at energies up to hundreds of
keV. However, at energies of the order of a gigaelectronvolt and
higher, the binding energy of electrons in atoms is significantly
manifested only in the small lower parts of the regions of
transmitted momentum and energies, which makes it necessary to
consider the transmission of momentum to electrons of all atomic
subshells which populations $f_i$  \cite{sco} satisfy the
condition
\begin {equation} 
Z - Z_v = \sum_{i=1}^{n_{sh}} f_i.
\end{equation}
In the theory of X-ray scattering  \cite{hub}, the contribution of
all atomic shells is described by the incoherent scattering
function
\begin {equation} 
\begin{array}{c}
S(k) = \sum_{n\neq 0}\langle 0|\sum_a e^{i\bm k \bm r_a} | n
\rangle \langle n|\sum_b e^{-i\bm k \bm r_b} | 0 \rangle \\
= Z + \langle 0|\sum_{a \neq b} e^{i\bm k (\bm r_a - \bm r_b)} | 0
\rangle -|F(k)|^2
\end{array}
\end{equation}
in which the summation is carried out over the excited atomic
states $n$ and the radius-vectors of the atomic electrons  $a$,
$b$, and
\begin {equation} 
F(k) = \langle 0|\sum_a e^{i\bm k \bm r_a} | 0 \rangle
\end{equation}
is atomic form factor. These values obey condition
\begin {equation} 
S(k \rightarrow 0) \simeq k^2 \bigg\langle \bigg( \sum_a x_a
\bigg)^2 \bigg\rangle = k^2 \sum_a \bigg\langle x_a^2
\bigg\rangle, \hspace{1cm} S(k \rightarrow \infty) = Z,
\end{equation}
\begin {equation} 
F(k \rightarrow 0) \simeq Z - \frac{1}{2} k^2 \bigg\langle \bigg(
\sum_a x_a \bigg)^2 \bigg\rangle, \hspace{1cm} F(k \rightarrow
\infty) = 0.
\end{equation}
Numerical dependence of the scattering function and form factor is
given in tables \cite{hub}. The use of scattering functions to
describe incoherent electron scattering was proposed by U. Fano
\cite{fan}. However, to demonstrate a qualitative picture in the
logarithmic approximation, it is sufficient to use a simplified
representation of the dynamic form factor
\begin {equation} 
S(\bm k, \omega) = \frac{k^2}{4 \pi^2 e^2} \: \Im
\frac{-1}{\varepsilon(\bm k, \omega)} = \sum_{i=1}^{n_{sh}}
\frac{f_i k^2}{k^2+2 m E_i} \delta \bigg( \omega - E_i -
\frac{k^2}{2 m} \bigg)
\end{equation}
obtained by substituting into (15) the approximation \cite{rh, da,
med}
\begin {equation} 
\Im \frac{-1}{\varepsilon(\bm k, \omega)} =
\frac{\pi}{2}\sum_{i=1}^{n_{sh}} \frac{\omega_{pi}^2}{\omega}
\delta \bigg( \omega - E_i - \frac{k^2}{2 m} \bigg),
\end{equation}
where $E_i$ is the ionization energy of the i-th atomic subshell
and
\begin {equation} 
\omega_{pi}^2 = \frac{4 \pi e^2 N f_i}{m}
\end{equation}
is plasma frequency calculated for the number density $N f_i$ of
electrons of the i-th electronic subshell. The parameters of the
Drude-Lorentz model oscillators $E_i$ and $f_i$ are found by
fitting the spectrum of the optical loss function, different
versions of which still lead to different results \cite{rh, da,
med}. As a simple approximation for $E_i$ and $f_i$, one can use
the ionization potentials and populations of the atomic subshells,
as suggested in \cite{bet}, and assume that the transmitted energy
consists of the value of the ionization potential and the kinetic
energy  $k^2/2m$ of the released electron
\begin {equation} 
\omega_{n0} \rightarrow E_i + \frac{k^2}{2m}~.
\end{equation}
At that the role of summation over the energy of the final atomic
state is played by integration over the electron momentum. Further
refinement needs both atomic structure calculations \cite{hub} and
experimental data \cite{pdg}.

Let us recall that in the X-ray range, the permittivity is given
by the expression $\varepsilon(\omega) \simeq 1 -
\omega^2_p/\omega^2$, which includes the plasma frequency
calculated only for electrons of the shells, whose binding energy
is lower than the excitation energy $\omega$. In this case, the
square of the transmitted 4-momentum in the denominator of (14)
takes the form
\begin {equation} 
q^2 = k^2 - \varepsilon(\omega_{n0})\omega_{n0}^2 \simeq k^2_\perp
+ \omega_{n0}^2/v^2 \gamma^2 + \omega_p^2 \simeq k^2_\perp +
\omega_p^2~,
\end{equation}
demonstrating that the influence of refraction allows one to
neglect the contribution of the atomic transition frequency
$\omega_{n0}$. Substituting (24) and (15) into (14) leads to the
expression
\begin {equation} 
\displaystyle{\bigg\langle \frac{d
p^2_\perp(k_1)}{dz}\bigg\rangle_{sh} = \frac{4 \pi^2 e^2 N}{v^2}
\int_0^{k_1} \frac{k^2_\perp}{(k^2_\perp + \omega_{n0}^2/v^2
\gamma^2 + \omega_p^2 )^2} \sum_{i=1}^{n_{sh}} \frac{f_i
k^2}{k^2+2 m E_i} d^2 \: k_\perp}~.
\end{equation}
Before continuing our consideration of the latter, let us dwell on
the role of the current contribution in (14), which, as is known
 \cite{ll4}, is noticeable in the theory of stopping power of
ultrarelativistic particles. This contribution arises at the lower
limit of integration. Indeed, for "small" transmitted momenta $k <
k_0 \sim \sqrt{2 m \omega_{n0}}$, $\omega_{n0} \ll \sqrt{2 m
\omega_{n0}}$, the exponential of the matrix element (14) can be
expanded in a series
\begin {equation} 
\langle n|e^{i\bm k \bm v}-i \bm v ( \bm\nabla e^{i\bm k \bm v} -
e^{i\bm k \bm v} \bm\nabla )/2m| 0 \rangle \simeq i(\bm k +
\omega_{n0} \bm v) \langle n|\bm r| 0 \rangle.
\end{equation}
By integrating the square of the (30) modulus over the azimuthal
angle \cite{ll4}, we are convinced that the main part of the
current contribution is cancelled by that of the longitudinal
momentum component in the expansion of the first term of (30),
leaving only the term, relativistically suppressed compared to the
contribution of transverse momentum component:
\begin {equation} 
\int_0^{2 \pi }|(\bm k + \omega_{n0} \bm v) \langle n|\bm r| 0
\rangle |^2 d\varphi = 2 \pi |x_{n0}|^2 \bigg(k^2_\perp +
\frac{\omega_{n0}^2}{v^2 \gamma^2 }  -
\frac{\omega_{n0}^2}{\gamma^2 }\bigg) = 2 \pi |x_{n0}|^2
\bigg(k^2_\perp + \frac{\omega_{n0}^2}{v^2 \gamma^4 } \bigg).
 \end{equation}
 The calculation of the mean square scattering
angle differs from the calculation of stopping power in that in
the latter case, instead of the square of the transverse component
of the transmitted momentum $k_\perp^2$, the energy of atomic
excitation $\omega_{n0}$ appears under the integral over the
transmitted momentum:
\begin {equation} 
\displaystyle{\bigg\langle -\frac{d \varepsilon}{dz}\bigg\rangle_{
sh} = \frac{4 \pi^2 e^2 N}{v^2} \int_0^{k_1}
\frac{k^2_\perp}{(k^2_\perp + \omega_{n0}^2/v^2 \gamma^2 +
\omega_p^2 )^2} \sum_{i=1}^{n_{sh}} \frac{f_i \omega_{n0}}{k^2+2 m
E_i} d^2 \: k_\perp}.
\end{equation}
Let us show that since, unlike $k^2_\perp$, the atomic excitation
energy $\omega_{n0}$ does not tend to zero together with
$k_\perp$, the contributions of the lower limit to integrals (29)
and (32) are essentially different.

We will begin our consideration with the limit of low medium
density $\omega_p^2 \ll \omega_{n0}^2/v^2 \gamma^2$. Since there
are no collective excitations of the medium in this case, the
contribution of the plasma frequency in the denominator of (29)
and (32) does not appear, and the summation is carried out over
all atomic shells:
\begin {equation} 
\begin{array}{l}
\displaystyle{\bigg\langle \frac{d
p^2_\perp(k_1)}{dz}\bigg\rangle_{sh}  \simeq \frac{4 \pi^2 e^2
N}{v^2} \bigg[Z ln\frac{k_1^2}{2mI} - \frac{1}{\gamma^2 v^2 m}
\sum_{i=1}^{n_{sh}} f_i E_i  \bigg(ln \frac{2 \gamma^2 v^2 m}{E_i}
-\frac{1}{2} \bigg) \bigg] } \\
\displaystyle{\simeq \frac{4 \pi^2 e^2 N Z}{v^2} \:
ln\frac{k_1^2}{2mI}},
\end{array}
\end{equation}

\begin {equation} 
\displaystyle{\bigg\langle -\frac{d \varepsilon
(k_1)}{dz}\bigg\rangle_{ sh} = \frac{2 \pi e^4 N Z}{m v^2} \bigg(2
ln \frac{k_1 v \gamma}{I} -1 \bigg) },
\end{equation}
where
\begin {equation} 
ln I = \frac{1}{Z}\sum_{i=1}^{n_{sh}}E_i
\end{equation}
is an approximate analogue of the mean excitation energy in the
theory \cite{bet, ll3, ll4, ll8}. Comparison of expressions (33)
and (34) shows that only in the case of stopping power evaluation
at the lower limit of integration does the equal to minus one
contribution comparable to the leading logarithmic term arise.
Furthermore, we note that expressions (33), (34) are related by a
factor $2m$, which differs from the factors used in \cite{bir,
st}.

Let us now consider the opposite limit $\omega_p^2 \gg
\omega_{n0}^2/v^2 \gamma^2$, realized at highly ultrarelativistic
energies of particles in crystals, in which the electrons of the
outer shells of metal and semiconductor atoms are collectivized
and should not be taken into account in the summation in Eqs. (29)
and (32). Moreover, in expression (28) for the square of the
transmitted 4-momentum $q$, the contribution of the excitation
energies of the inner shells $\omega_{n0}$  can be neglected and,
similarly to (34), for the contribution of these shells, we obtain
from (32)
\begin {equation} 
\displaystyle{\bigg\langle -\frac{d \varepsilon
(k_1)}{dz}\bigg\rangle_{sh} \simeq \frac{2 \pi e^4 N }{m v^2} (Z -
Z_v) \bigg(2 ln \frac{k_1}{\omega_p} -1 \bigg) }.
\end{equation}
Taking into account equations (15)-(17), the contribution of the
excitation of valence electrons to stopping power is equal to
\begin {equation} 
\begin{array}{l}
\displaystyle{\bigg\langle -\frac{d \varepsilon
(k_1)}{dz}\bigg\rangle_v = \frac{4e^4}{v^2} \int_0^{k<k_1}d^2
k_\perp \int_0^\infty \omega d \omega \sum_{n\neq 0, N
Z}\frac{\delta(\omega-\omega_{n0})}{|k^2_\perp
+\omega^2/v^2-\varepsilon(\bm k, \omega)\omega^2|^2 }
\frac{|\langle n|e^{i\bm k \bm v}| 0 \rangle|^2}{|\varepsilon(\bm
k, \omega)|^2}}\\
\displaystyle{= \frac{4e^4}{v^2} \int_0^{k<k_1}d^2 k_\perp
\int_0^\infty \omega d \omega \frac{S(\bm k, \omega)}{|k^2_\perp
+\omega^2/v^2-\varepsilon(\bm k, \omega)\omega^2|^2 } }
\\
\displaystyle{= \frac{2 \pi e^4 n_v}{m v^2} \int_0^{k<k_1}
\frac{k^2 dk^2_\perp}{|k^2_\perp +\omega^2_{\bm
k}/v^2-\varepsilon(\bm k, \omega_{\bm k})\omega^2_{\bm k}|^2 } }\\
\displaystyle{\simeq \frac{2 \pi e^4 n_v }{m v^2}\bigg(ln
\frac{k_1^2}{\omega_p^2 +E_g^2} -1 \bigg)\simeq \frac{2 \pi e^4
n_v }{m v^2}\bigg(2 ln \frac{k_1}{\omega_p} -1 \bigg) },
\end{array}
\end{equation}
where the equal to $\omega_p$ choice of the effective cutoff
momentum can be accepted as consistent with the accuracy of the
logarithmic approximation. Finally, summing (36) and (37), we
obtain
\begin {equation} 
\displaystyle{\bigg\langle -\frac{d \varepsilon
(k_1)}{dz}\bigg\rangle = \bigg\langle -\frac{d \varepsilon
(k_1)}{dz}\bigg\rangle_v + \bigg\langle -\frac{d \varepsilon
(k_1)}{dz}\bigg\rangle_{ sh} \simeq \frac{2 \pi e^4 N Z}{m v^2}
\bigg(2 ln \frac{k_1}{\omega_p} -1 \bigg) },
\end{equation}
where $NZ$ is the total electron concentration of the crystal.

Thus, we have verified that, when calculating stopping power, the
contribution of the lower limit of integration over the
transmitted momentum (the subtraction of unity in (34), (36) and
(38)) turns out to be not negligible compared to the leading
logarithmic term. At the same time calculation of the mean square
scattering angle in the low-density limit (33) showed that
replacing the excitation energy in the integrand of (32) with the
square of the transverse component of the transmitted momentum
leads to the contribution of the lower limit of integration
becoming negligibly small. It was important for this consideration
that the contribution of the longitudinal component of transmitted
momentum to the numerator of expression (8), which does not vanish
together with the transverse component of the momentum, is
cancelled, when the current term is taken into account \cite{ll4}.
Similarly to Eq. (33), the contribution of the lower limit of
integration does not lead to the appearance of an additional term
either when calculating the mean square projected scattering
angles in the limit $\omega_p^2 \gg \omega_{n0}^2/v^2 \gamma^2$
using expressions (18) and (29):
\begin {equation} 
\begin{array}{l}
\displaystyle{\bigg\langle \frac{d \theta^2_{x,y}}{dz}\bigg\rangle
= \frac{1}{2 p^2}\bigg\langle \frac{d
p^2_\perp(k_1)}{dz}\bigg\rangle_{g=0} } \\
\displaystyle{= \frac{2\pi e^4 N}{v^2 p^2} \bigg(Z_v
ln\frac{k_1^2}{2 m I_v} + \sum_{i=1}^{n_{sh}} f_i ln\frac{k_1^2}{2
m E_i} \bigg) } \\
\displaystyle{= \frac{2\pi e^4 n}{v^2 p^2} \: ln\frac{k_1^2}{2 m
I_v} }, \hspace{1 cm} ln I = \frac{1}{Z}\bigg(Z_v lnI_v +
\sum_{i=1}^{n_{sh}} f_i ln E_i \bigg),
\end{array}
\end{equation}
where the effective ionization energy $I$ is introduced, being
equal to 93 eV for silicon. As already noted, the plasma frequency
in (29), strictly speaking, should be calculated taking into
account the dependence of the effective electron concentration on
the excitation energy. However, given the condition $\omega^2_p
\ll 2 m E_i$, its contribution turns out to be insignificant,
making its strict definition unnecessary.

Thus, we have seen that, similar to stopping power theory
\cite{bet, ll3}, the need to account for the complex behavior of
permittivity and excitation of both collective electrons and
electrons of inner atomic shells makes the use of the logarithmic
approximation inevitable when calculating the mean square
scattering angle by electrons. In this case, the entire influence
of the substance structure is accounted for by a single parameter
of effective ionization energy $I$ while the logarithmic
dependence on the integration limit $k_1$ over the transferred
momentum allows the use of the resulting expression (39) to be
consistent with the description of scattering at large values of
the transmitted momentum. Moreover, the increase in the upper
limit of the latter with increasing particle energy reduces the
relative uncertainty associated with the use of the logarithmic
approximation.

\subsection{Scattering under relativistic momentum transmission to free
electrons}

The influence of electron binding has been considered in the
non-relativistic limit $\sqrt{2mI}\ll k_1 \ll mc$.  However, under
the transmeiited momenta $k \gg k_1$, the electron binding can be
neglected, when calculating the mean square angle of incoherent
scattering of particles of energy $\varepsilon$ and mass $M$ ($M =
m$ in the case of electrons and positrons), and atom electrons can
be considered free and at rest \cite{ll4}. We will represent the
formulas for the case of strongly ultrarelativistic particles with
Lorentz factors $\gamma = \varepsilon/M \gg 1$  in terms
\cite{ll4} of the 4-momentum transmitted to the electron
\begin {equation} 
q_i = \big(\sqrt{k^2 + m^2}-m, \bm k \big),
\end{equation}
the square of which is connected with the kinetic energy,
transmitted to the electron,
\begin {equation} 
T = \sqrt{k^2 + m^2}-m
\end{equation}
and its dimensionless value $\Delta = T/m$ by the relation
\begin {equation} 
-q^2 = 2mT = 2m^2 \Delta~.
\end{equation}
To establish how these quantities are related to the transverse
momentum transmission, we use the expression for the cosine of the
recoil electron emission angle  \cite{ll2}
\begin {equation} 
cos \theta = \frac{(\varepsilon+M)}{p}\frac{\sqrt{k^2 +
m^2}-m}{k},
\end{equation}
 allowing to obtain the ratio
\begin {equation} 
\frac{k^2_\perp}{2m} = \frac{k^2 sin^2 \theta}{2m} = T,\hspace{0.5
cm} T\ll T_{max}~,
\end{equation}
where
\begin {equation} 
T_{max} = m \Delta_{max} = \frac{2mp^2}{m^2 + M^2 + 2 m
\varepsilon}
\end{equation}
is the maximum energy transmitted to an electron during a
collision \cite{ll4, ll2}. From this relation it follows that the
mean square of the scattering angle of particles on free electrons
with limited energy transmissions is related to stopping power by
the factor $2m/p^2$ , and not by the factor $m/p^2$ used in
\cite{st}.

Relationship (44), however, breaks down when the energy
transmission (42) becomes comparable to the maximum transmitted
energy (45). In this case, from (43) the relation follows
\begin {equation} 
\frac{k^2_\perp}{2m} = \frac{T_{max} -T} {T_{max}}T,\hspace{0.5
cm} T \sim T_{max}~,
\end{equation}
which differs significantly from the relation (44) and reflects a
decrease in the scattering angle as the electron energy tends to
its maximum, in accordance with the fact that the latter is
achieved when the electron is emitted at a zero angle
corresponding to zero transverse momentum. Thus, relation (46)
also does not confirm the method of using the stopping power
formula in \cite{st} to describe multiple scattering.

Scattering cross section of a particle with mass $M \gg m$ on a
free electron \cite{ll4}
\begin {equation} 
d\sigma = \frac{2 \pi e^4}{v^2}\bigg(1- \frac{v^2
\Delta}{\Delta_{max}} \bigg) \frac{d\Delta}{\Delta^2},
\end{equation}
allows us to obtain for the contribution to the effective braking
\cite{ll4} of the energy interval $\Delta_{min} < \Delta <
\Delta_{max} $, where, see (28),
\begin {equation} 
\Delta_{min} = \omega_p^2/2 m^2,
\end{equation}
 an expression
\begin {equation} 
\int_{\Delta_{min}}^{\Delta_{max}} m \Delta\sigma = \frac{2 \pi
e^4}{m v^2}\bigg(ln \frac{\Delta_{max}}{\Delta_{min}} - v^2
\bigg),
\end{equation}
in which the square of the velocity of the original particle is
subtracted from the leading logarithmic term.

It should be clarified why such a term should not be considered
either in the calculation of the mean square scattering angle or
in that of the stopping power of channeled particles. The point is
that the transmission of momentum exceeding its maximum value for
the channeled particles
\begin {equation} 
p_{ch} = p \: \theta_{ch} = \sqrt{2 V_0 \varepsilon},
\end{equation}
leads to instantaneous dechanneling, accompanying single
scattering at an angle of order or greater than the channeling
angle. Therefore, when calculating the stopping power of channeled
particles, the transmitted momentum should be limited by the value
(50), and the dimensionless transmitted energy - by
\begin {equation} 
\Delta_{ch} = p^2_{ch}/2 m^2 = V_0 \varepsilon/m^2,
\end{equation}
leading to the expression
\begin {equation} 
\int_{\Delta_{min}}^{\Delta_{ch}} m \Delta\sigma = \frac{2 \pi
e^4}{m v^2}\bigg(ln \frac{\Delta_{ch}}{\Delta_{min}} -
\frac{\Delta_{ch}}{\Delta_{min}} \bigg),
\end{equation}
the correction to the logarithm in which is only
\begin {equation} 
\frac{\Delta_{ch}}{\Delta_{min}} \simeq \frac{V_0 (M^2 + 2 m
\varepsilon)}{2 m^2 \varepsilon} \sim 10^{-4}.
\end{equation}
We also note that the ratio of the integral (52), describing the
process of diffuse scattering of channeling particles, to the
integral (49), which additionally includes the contribution of
catastrophic dechanneling processes, as in \cite{st}, is close to
0.60 in the examples we will consider below for illustration. The
resulting overestimation of the mean square of the electron
scattering angle was phenomenologically compensated in \cite{st}
by multiplying it by 0.5. However, in fact, an adequate
consideration of the energy transmissions $\Delta_{ch}< \Delta <
\Delta_{max}$ is its modeling based on the single scattering cross
section. Similar conclusions are obtained when considering the
incoherent scattering of electrons and positrons ($M = m$), for
which the subtracted value in the brackets of formula (47) should
be multiplied by 2 and the value v$\Delta_{max}$ set equal to $
\gamma$.

The use of formulas (47) and (52) to describe electron scattering
under conditions of non-uniform averaged electron distribution in
crystals implies the use of a local approximation in which the
electron number density is taken to be equal to its averaged value
on the particle trajectory. This approximation does not take into
account that momentum and energy exchange occurs over transverse
distances, ranging from those much smaller than interatomic
spacing to those comparable to and exceeding the latter. Both the
electron number densiy in a crystal and their binding energy
distribution at such distances differ significantly from the
corresponding local characteristics on the particle's trajectory.
A quantitative description of incoherent scattering, taking these
factors into account, is the primary goal of this work. To address
the latter, an expression for the mean square of the electron
scattering angle in crystals will be derived in the next section.
As noted above, this expression should include only the
contribution of transverse momenta not exceeding the maximum
momentum of the channeled particles (50), while the process of
larger transmissions should be considered as single scattering,
typically leading to catastrophic dechanneling. To ensure the
correct use of the simplest local approximation based on the
Coulomb cross section and the local average electron density along
the particle's trajectory, the reciprocal of the maximum momentum
of the channeled particles, $1/p_{ch}$, ??should not exceed the
typical distance of change in the average electron density, which
in the region of stable channeling of positively charged particles
should be estimated as a few hundredths of an angstrom, limiting
the energy of channeled positrons from below by several hundred
MeV.

\section{Integration over the transmitted momentum in the case of a
crystal}

The contribution of nonzero reciprocal lattice vectors to the
averaged product of the transverse components of the incoherent
scattering momentum can be written in the form
\begin {equation} 
\bigg\langle \frac{d p_i p_j (\bm b)}{dz}\bigg\rangle = \frac{1}{4
\pi^2 v^2 \Omega} \sum_{\bm g} S_{cr}(\bm g) e^{-g^2 u_1^2/2}e^{i
\bm g \: \bm b} \int d^2 k_\perp \frac{k_{+i}k_{-j} S(\bm k_+, \bm
k_-)}{(k^2_+ + \omega^2_p)(k^2_- + \omega^2_p)}~,
\end{equation}
where the incoherent scattering function of two vector arguments
\begin {equation} 
S(\bm k_+, \bm k_-) = \sum_{n\neq 0}\langle 0|\sum_a e^{i\bm k_-
\bm r_a} | n \rangle \langle n|\sum_b e^{-i\bm k_+ \bm r_b} | 0
\rangle
\end{equation}
is introduced. Let us explain why, in contrast to (14), the
permittivity is taken into account here based on the asymptotic
formula $\varepsilon(\omega) \simeq 1 - \omega^2_p/\omega^2$. A
similar approach could not be used for g = 0 due to the quadratic
in $k^2_\perp$ dependence of the denominator $|k^2_\perp
+\omega^2/v^2-\varepsilon(\bm k, \omega)\omega^2|^2$ at small
transmitted momenta, leading to a strong dependence on the
permittivity, which differs significantly from unity at low
frequencies. However, in the case of $\bm g \neq 0$, the
dependence of the denominator of (54) on the square of the
momentum near one of the minima, for example, the minimum at $ k_-
\rightarrow 0$, becomes linear $(k^2_+ + \omega^2_p)(k^2_- +
\omega^2_p) \simeq (g^2 + \omega^2_p)(k^2_- + \omega^2_p)$, as a
result of which the estimate of the contribution of the
integration region $k_- < k_0 \ll g$ takes the form
\begin {equation} 
\int_{k<k_0} d^2 k_\perp \frac{k_{+i}k_{-j}\: S(\bm k_+, \bm
k_-)}{(k^2_+ + \omega^2_p)(k^2_- + \omega^2_p)} \simeq
\frac{k_0^2}{g^2 + \omega^2_p} \bigg[1- \frac{\omega^2_p}{k_0^2}
\:ln \bigg(\frac{k_0^2}{\omega^2_p}+1 \bigg) \bigg] \simeq
\frac{k_0^2}{g^2},
\end{equation}
demonstrating the possibility of neglecting the contribution of
the plasma frequency at $k_1 \gg k_0 \gg \omega_{pl}$ which
eliminates the need to take into account more precise frequency
dependence of the permittivity.

A complete calculation of the incoherent scattering function of
two vector arguments (55) has not yet been carried out, and the
main attention in the works \cite{esb, bur, esb2} is given to the
representation
\begin {equation} 
S(\bm k_+, \bm k_-) = F(g)+\sum_{a\neq b}\langle 0|e^{i\bm k_- \bm
r_a -i\bm k_+ \bm r_b} | 0 \rangle - F(k_+)F(k_+)
\end{equation}
 was paid to the form factor $F(g)$ of electron density. The values of the
latter can be found in tables \cite{hub} or calculated using
formulas applicable for large momenta \cite{tob, idf}, the list of
which does not include the Doyle-Turner approximation used in
[16]. With a further increase in the accuracy of the whole model,
it is also possible to take into account small differences in the
form factor values given in \cite{hub} from that measured in the
crystal \cite{lu}. In the absence of results of a numerical
calculation of the scattering function of two vector arguments
$S(\bm k_+, \bm k_-)$, to estimate the second term on the
right-hand side of (57) we will use the ansatz
\begin {equation} 
\sum_{a\neq b}\langle 0|e^{i\bm k_- \bm r_a -i\bm k_+ \bm r_b} | 0
\rangle \rightarrow \sqrt{(S(k_+)+F^2(k_+)-Z)(S(k_-)+F^2(k_-)-Z)},
\end{equation}
suggested by the representation of the "ordinary" scattering
function of one argument (20). The absence of dependence on the
variable $\bm k$ of the first term of (57) allows us to use the
integrals
\begin {equation} 
\int_{k<k_1} \frac{k^2_\parallel -g^2/4}{k^2_- k^2_+} \frac{d^2
k}{\pi}= \frac{1}{2}\bigg[ln \bigg(\frac{k_1^2}{g^2}+\frac{1}{4}
\bigg) -1\bigg] \simeq ln \frac{k_1}{g} -\frac{1}{2},
\end{equation}
\begin {equation} 
\int_{k<k_1} \frac{k^2_\perp}{k^2_- k^2_+} \frac{d^2 k}{\pi}=
\frac{1}{2}\bigg[ln \bigg(\frac{k_1^2}{g^2}+\frac{1}{4} \bigg) +
1\bigg] \simeq ln \frac{k_1}{g} + \frac{1}{2}
\end{equation}
and obtain in the limit $k_1 \gg  g$
\begin {equation} 
\begin{array}{l}
\displaystyle{\int_{k<k_1} \frac{S(\bm k_+, \bm k_-)
(k^2_\parallel -g^2/4)}{(k^2 -g^2/4)^2 - (\bm k \bm g)^2}
 \frac{d^2
k}{\pi}= I_{1,p}(g,k_1)+I_{2,p}(g,k_1)}\\
\displaystyle ={F(g)\bigg(ln \frac{k_1}{g} -\frac{1}{2} \bigg) +
\int_{k<k_1} \frac{[S(\bm k_+, \bm k_-)-F(g)](k^2_\parallel
-g^2/4) }{(k^2 -g^2/4)^2 - (\bm k \bm g)^2} \frac{d^2 k}{\pi}},
\end{array}
\end{equation}
\begin {equation} 
\begin{array}{l}
\displaystyle{\int_{k<k_1} \frac{S(\bm k_+, \bm
k_-)k^2_\perp}{(k^2 -g^2/4)^2 - (\bm k \bm g)^2}
 \frac{d^2
k}{\pi}= I_{1,n}(g,k_1)+I_{2,n}(g,k_1)}\\
\displaystyle ={F(g)\bigg(ln \frac{k_1}{g} +\frac{1}{2} \bigg) +
\int_{k<k_1} \frac{[S(\bm k_+, \bm k_-)-F(g)]k^2_\perp }{(k^2
-g^2/4)^2 - (\bm k \bm g)^2} \frac{d^2 k}{\pi}}.
\end{array}
\end{equation}

\begin{figure} 
\label{Fig1}
 \begin{center}
\resizebox{80mm}{!}{\includegraphics{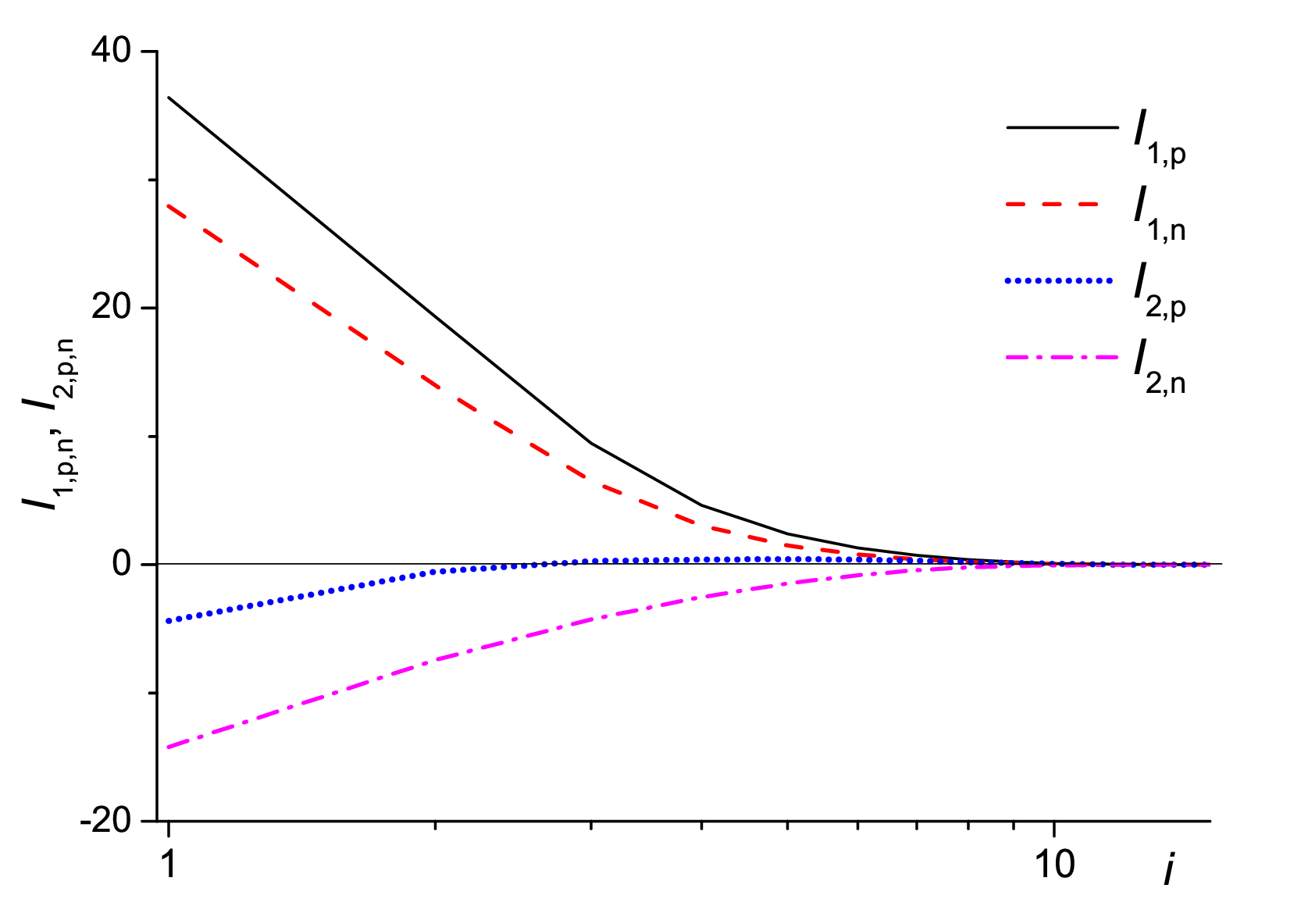}}
\caption{Dependence on the index $i = 1, 2,..$ of the treciprocal
lattice vectors $q_i = 2 \pi i/d$ of the contributions of the
first and second terms in expressions (61), (62) for scattering in
planes parallel (p) and normal (n) to this vector for the case of
the (110) Si plane and boundary momentum (50) evaluated at 2 GeV
positron energy.}
\end{center}
\end{figure}

As we can see, the contribution of each reciprocal lattice vector
$\bm g$ is characterized by the projected mean square scattering
angles in the parallel (p), (61), and normal (n), (62), scattering
planes. In the axial case, the reciprocal lattice vectors $\bm g$
are noncollinear, and the description of scattering, taking into
account its asymmetry, must be based on determining, at each point
of the trajectory, the azimuthal orientation of a pair of
perpendicular planes in which the sum of these contributions
reaches its maximum and minimum values, as well as these values
themselves, using the sum of expression (54) and the contribution
(39) of scattering processes without momentum transfer to the
lattice. Note that, if we neglect the difference between the
integrals (61), (62), even at zero angle of incidence on some
crystal  axis, incoherent scattering will remain asymmetric due to
the influence of the effect of particle channeling along the
planes passing through it. Below we will consider in detail the
simpler and more relevant for most modern applications incoherent
electron scattering in the planar case.

We assume that the direction of motion of the channeled particles
is sufficiently distant from the directions of the atomic chains
forming the planes, neglecting coherent momentum transmission
parallel to the atomic planes and considering the set of
reciprocal lattice vectors to be one-dimensional and normal to the
crystal planes. The dependences of the contributions of the first
and second terms in expressions (61), (62) for scattering in
planes parallel (p) and normal (n) to the reciprocal lattice
vectors with moduli $q_i = 2 \pi i/d$, $i = 1, 2,..$, for the case
of the (110) Si plane with interplanar distance $d = a/2\sqrt{2} =
1.92 \AA$ are shown in Fig. 1. The boundary momentum $k_1$ is
chosen to be equal to the maximum momentum (50) of 2 GeV channeled
positrons, the value of which is sufficient to consider the
integral included in (61), (62) to be independent of the upper
limit. Note the noticeable asymmetry of the scattering and the
significance of the integral contribution, which were not
considered in \cite{esb, bur, esb2}.

Incoherent electron scattering in crystals is locally described by
the sum of contributions (39) and (54) of zero and non-zero
momentum transmissions to the crystal lattice:
\begin {equation} 
\begin{array}{l}
\displaystyle{\bigg\langle \frac{d \theta^2_{x}}{dz}\bigg\rangle =
\frac{4\pi e^4 N}{v^2 p^2} \left\{ Z ln \frac{k_1}{\sqrt{2 m I}}+
\sum_{i=1,2,..}S_{cr}(g_i) exp\bigg(i g_i
x - \frac{g_i^2 u_1^2}{2} \bigg) \right. }\\
 \displaystyle{\left. \times \bigg[ F(g_i)\bigg(ln
\frac{k_1}{g_i} -\frac{1}{2} \bigg) + \int_{k<k_1} \frac{[S(\bm
k_+, \bm k_-)-F(g_i)](k^2_x -g_i^2/4) }{(k^2 -g_i^2/4)^2 - (\bm k
\bm
g_i)^2} \frac{d^2 k_\perp}{\pi} \right\} }\\
=\displaystyle{ \frac{4\pi e^4 n(x)}{v^2 p^2} ln
\frac{k_1}{k_{eff,x}(x)}},
\end{array}
\end{equation}

\begin {equation} 
\begin{array}{l}
\displaystyle{\bigg\langle \frac{d \theta^2_{y}}{dz}\bigg\rangle =
\frac{4\pi e^4 N}{v^2 p^2} \left\{ Z ln \frac{k_1}{\sqrt{2 m I}}+
\sum_{i=1,2,..}S_{cr}(g_i) exp\bigg(i g_i
x - \frac{g_i^2 u_1^2}{2} \bigg) \right. }\\
 \displaystyle{\left. \times \bigg[ F(g_i)\bigg(ln
\frac{k_1}{g_i} +\frac{1}{2} \bigg) + \int_{k<k_1} \frac{[S(\bm
k_+, \bm k_-)-F(g_i)]k^2_y }{(k^2 -g_i^2/4)^2 - (\bm k \bm
g_i)^2} \frac{d^2 k_\perp}{\pi} \right\} }\\
=\displaystyle{ \frac{4\pi e^4 n(x)}{v^2 p^2} ln
\frac{k_1}{k_{eff,y}(x)}}.
\end{array}
\end{equation}
In the last lines of both expression (63) and (64) effective
incoherent scattering momenta $k_{eff,x}(x)$ and $k_{eff,y}(x)$
are introduced that take into account the azimuthal asymmetry for
planes parallel and perpendicular to the crystal plane.

\begin{figure} 
\label{Fig2}
 \begin{center}
\resizebox{55mm}{!}{\includegraphics{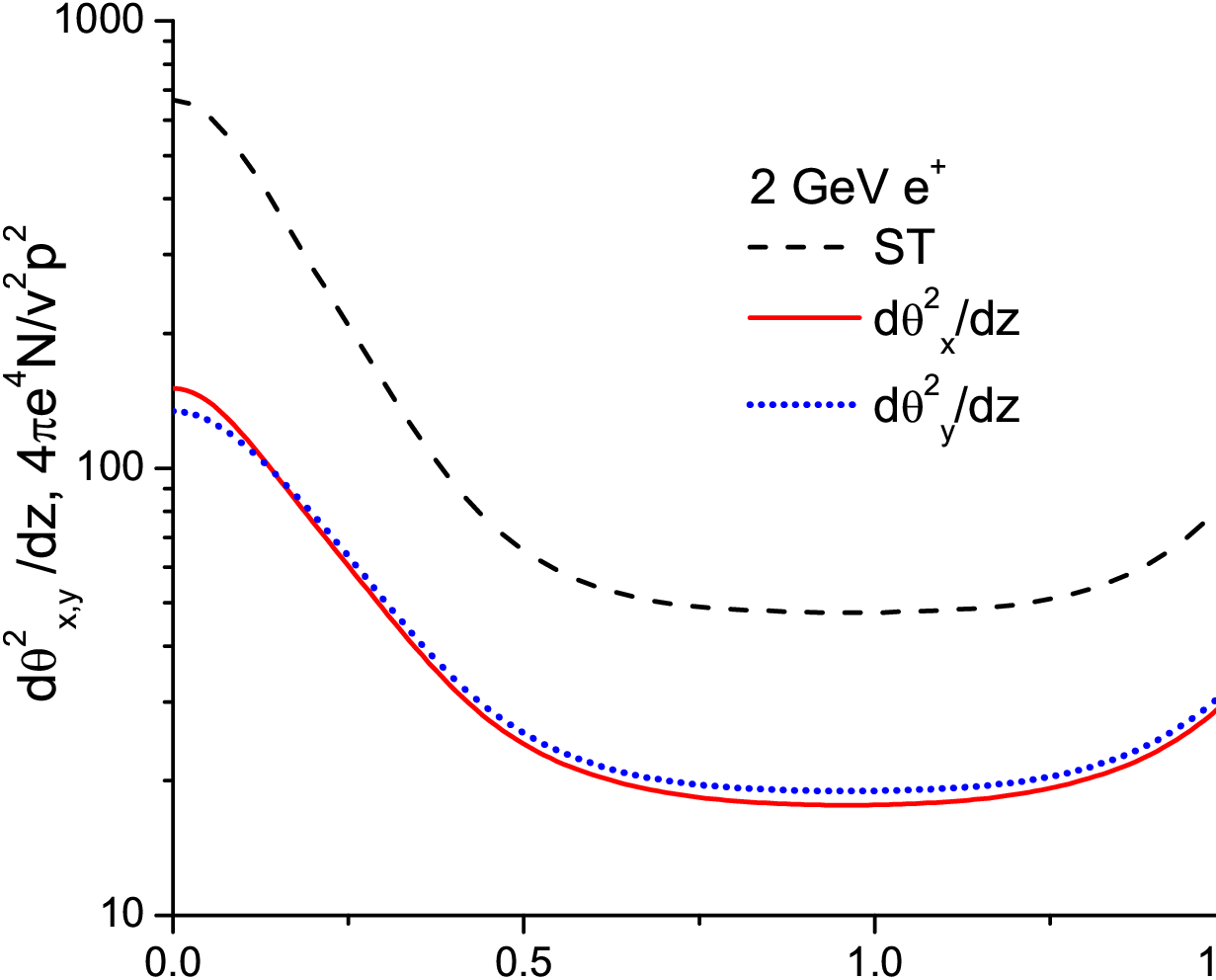}} \hspace{20mm}
\resizebox{55mm}{!}{\includegraphics{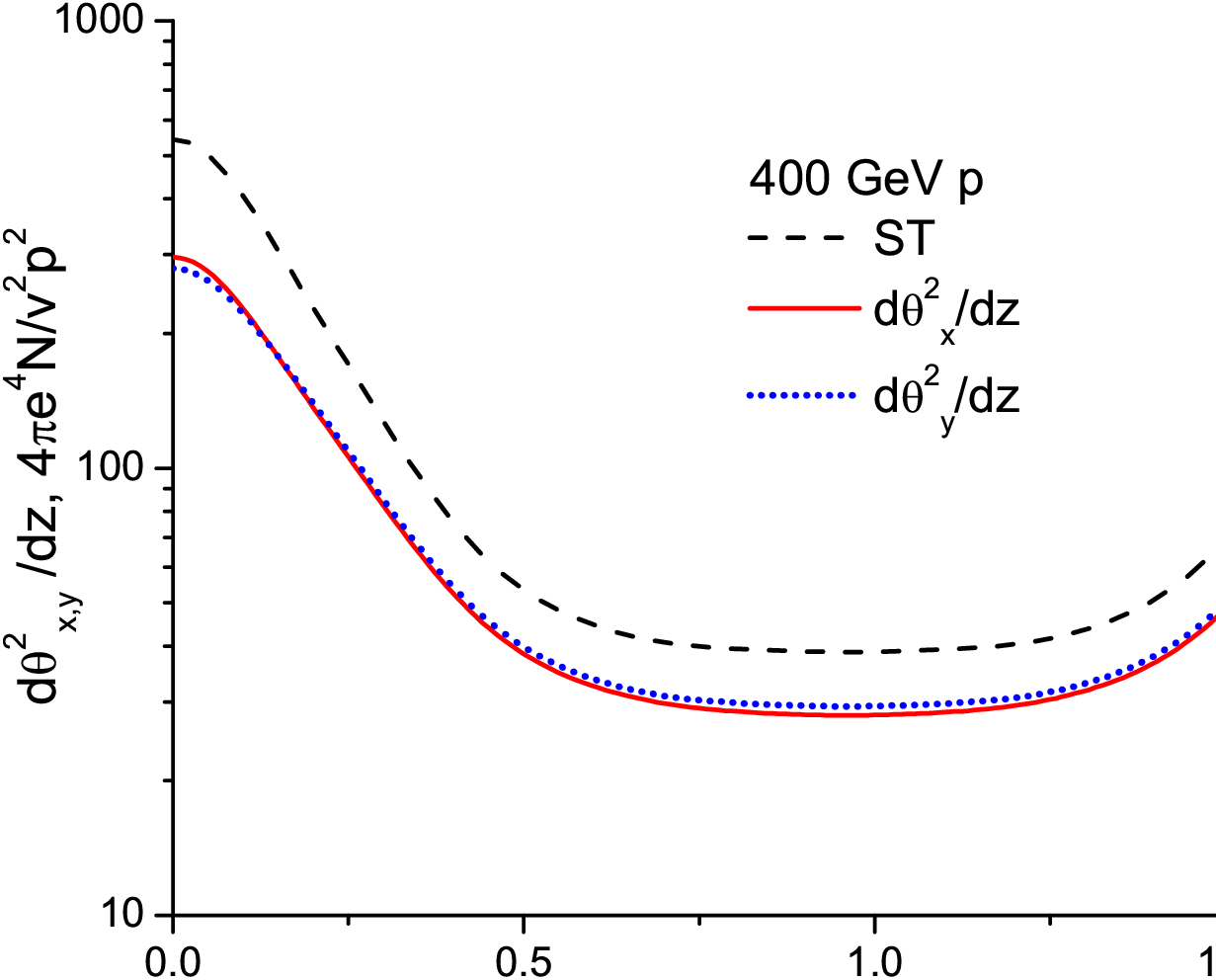}}\\
\caption{The mean squares of the incoherent scattering angles on
electrons per unit length in the xz and yx planes, calculated for
2 GeV positrons (left) and 400 GeV protons (right), as functions
of the distance to the (110) atomic plane of a silicon crystal.
These angles were calculated using formulas (63) and (64) (solid
and dotted curves) and formula (65) \cite{st} divided by two
(dashed curve). All quantities are expressed in units of the
multiplier in front of the curly brackets in formulas (63) and
(64).}
\end{center}
\end{figure}

In Fig. 2, the results of calculation using formulas (63), (64)
are compared with the prediction of the formula
\begin {equation} 
\frac{d \langle \theta^2 \rangle }{dz} = \frac{m}{p^2} \bigg(
-\frac{d\varepsilon}{dz}\bigg) \frac{n(x)}{N Z},
\end{equation}
used in \cite{st} for the nonprojected (space) (for the latter
definition see \cite{pdg}) mean square scattering angle over a
unit length, into which, at the energies under consideration, the
formula
\begin {equation} 
-\frac{d\varepsilon}{dz} = \frac{2 \pi e^4 N Z}{m} \bigg(ln
\frac{4 m^2 \gamma^2}{\omega^2_p} - 1 \bigg)
\end{equation}
for stopping power, corresponding to effective braking (49), which
takes into account the density effect \cite{ll8}, should be
substituted.

As examples, we consider the case of positrons with an energy of 2
GeV, optimal for use in crystalline undulators \cite{tik}, and the
case of protons with an energy of 400 GeV, extensively studied
experimentally in \cite{st}, which also illustrates the situations
of extracting high-energy beams from storage rings and the
experiment \cite{bar3, kim, fom, aio} being prepared to measure
the electromagnetic moments of short-lived particles. Since Fig. 2
illustrates scattering in the xz and yz planes, the results of the
calculation using formula (65) for the nonprojected (space) mean
square scattering angle are divided by two.

As was shown in Section III, the inclusion in formula (66) for
stopping power of a contribution from a spanning several orders of
magnitude range of energy, the loss of which results in
instantaneous dechanneling, leads to an overestimation of the
square of the scattering angle. To compensate for this, a
correction factor of 1/2 was used in formula (65) in \cite{st},
which is evidently inconsistent with the kinematic relation (42)
between energy and momentum. Fig. 2 shows that such a modification
of relation (65) indeed leads to an improvement in the agreement
with the calculation using formulas (63) and (64). However, at
positron energy of 2 GeV, while the Lorentz factor increases, the
maximum momentum (50) of the channeled positrons decreases, as a
result of which the relative fraction of energies, whose
transmission leads to instantaneous dechanneling, increases. As a
result, formula (65) gives a more overestimated prediction at 2
GeV, for the correction of which the factor $1/4$ from \cite{bir}
is more suitable.

Let's move on to a numerical comparison of the diffuse and
instantaneous dechanneling processes based on their local
characteristics and define the "diffuse" dechanneling length by
the relation
\begin {equation} 
\frac{p v}{2} \frac{d \langle \theta^2_{x}(x) \rangle }{dz}
l_{diff}= \frac{2 \pi e^4 n(x)}{v p}\: ln \bigg( \frac{p_{ch}
(x)}{k_{eff, x,y}}\bigg)l_{diff} = V_{max} -V(x),
\end{equation}
when using formula (63) and substitute the momentum of channeled
particles at the point under consideration $p_{ch} (x) =
\sqrt{2[V_{max} -V(x)]\varepsilon}$ as the boundary momentum
$k_1$.

\begin{figure} 
\label{Fig3}
 \begin{center}
\resizebox{55mm}{!}{\includegraphics{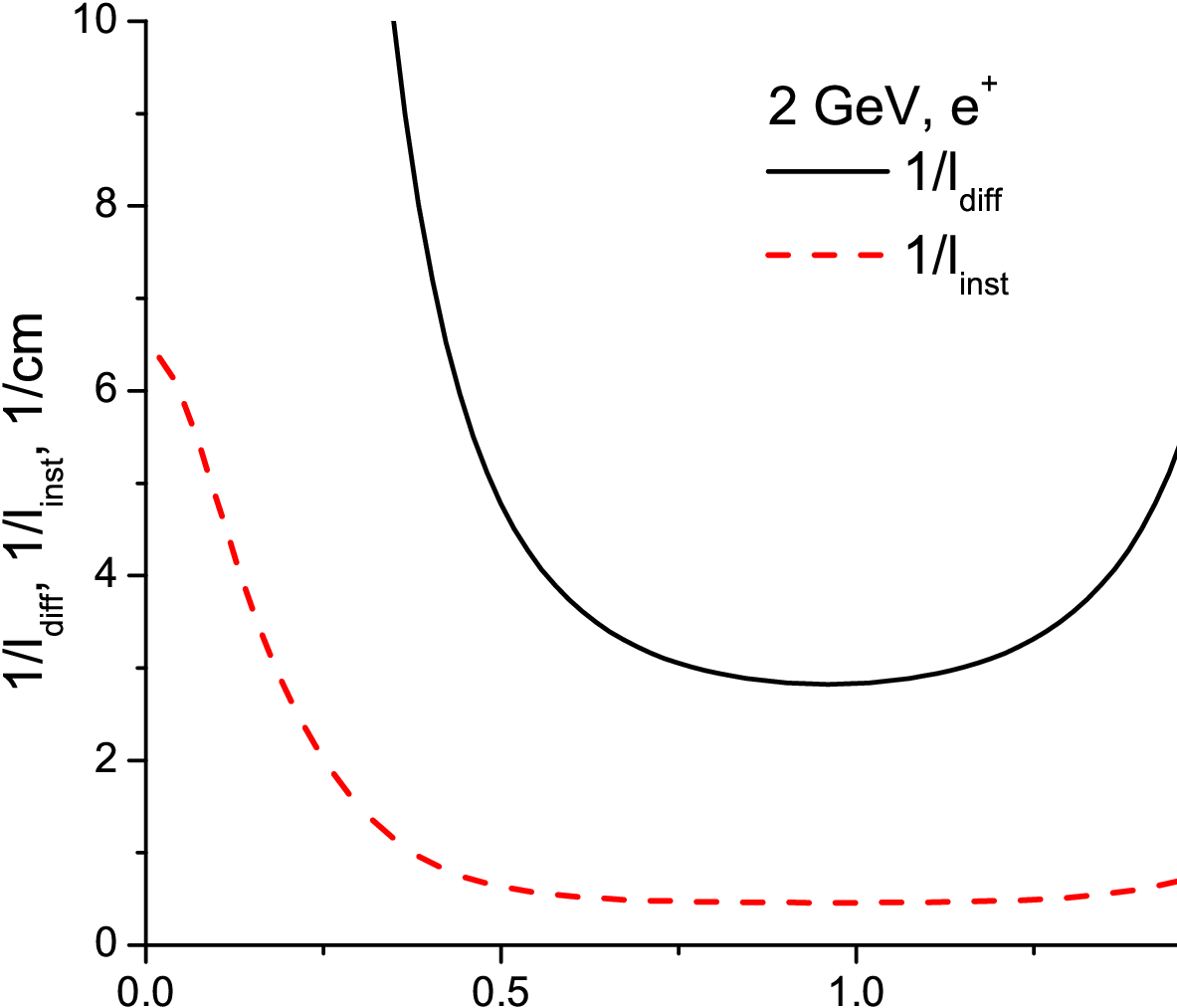}} \hspace{20mm}
\resizebox{55mm}{!}{\includegraphics{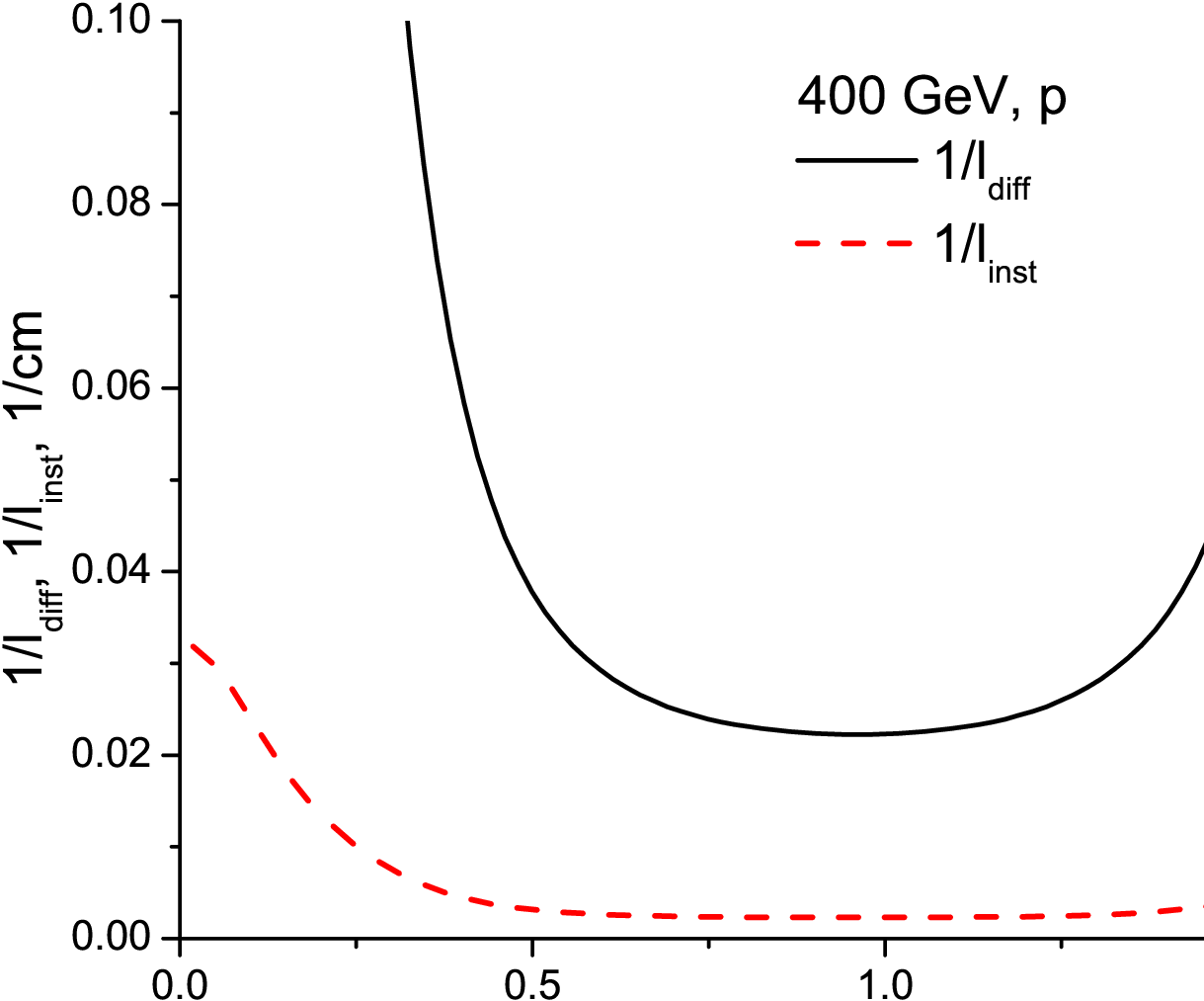}}\\
\caption{The reciprocal values of the "diffuse" and
"instantaneous" effective dechanneling lengths calculated for 2
GeV positrons (left) and 400 GeV protons (right) as a function of
the distance to the (110) atomic plane of a silicon crystal. }
\end{center}
\end{figure}
We assune the magnitude of the momentum transmission during
instantaneous dechanneling to be sufficiently high to use a local
approximation by introducing its "instantaneous" length using the
simple relation
\begin {equation} 
n(x)\int d\sigma \cdot l_{inst} = \frac{4 \pi e^4 n(x)}{v^2 p^2}
2\int_{\theta_{ch}}^\infty \int_{-\infty}^\infty \frac{d
\theta_y}{\theta_x^2 + \theta_y^2} \: l_{inst} = \frac{e^4 n(x)
l_{inst}}{v p \: [ V_{max} -V(x)]} = 1,
\end{equation}
where $\theta_{ch} = p_{ch}(x)/p$. The reciprocals of the
introduced lengths (67), (68) are compared in Fig. 3, which shows
that the efficiency of instantaneous dechanneling turns out to be
\begin {equation} 
\frac{l_{inst}}{l_{diff}} = 2 \: ln\bigg(\frac{p_{ch} (x)}{k_{eff,
x,y}(x)}\bigg) \sim 5 \div 10
\end{equation}
times lower than that of diffuse dechanneling. The present
refinement of methods for the theoretical description and modeling
of electron scattering is most important in the region of the most
stable channeling motion and vanishingly small nuclear number
density, in which the intensity of electron scattering exceeds
that of nuclear scattering. Using $V_{max}/3$ instead of $V_{max}$
in the estimates above, the last estimate can be interpreted as of
100 particles leaving the region of most stable channeling motion,
10-20\% do so instanteneously as a result of single scattering.

At a boundary momentum $k_1 > 20 \times 2 \pi/a$, the integral in
expressions (63) and (64) ceases to depend on $k_1$, allowing us
to introduce independent of it local effective minimal momenta
$k_{eff,x}(x)$, $k_{eff,y}(x)$ for scattering planes xz and yz.
The coordinate dependence of the latter is compared in Fig. 4 with
the characteristic ionization momentum (39) and ionization
momentum of collectivised electrons (18).
\begin{figure} 
\label{Fig4}
 \begin{center}
\resizebox{70mm}{!}{\includegraphics{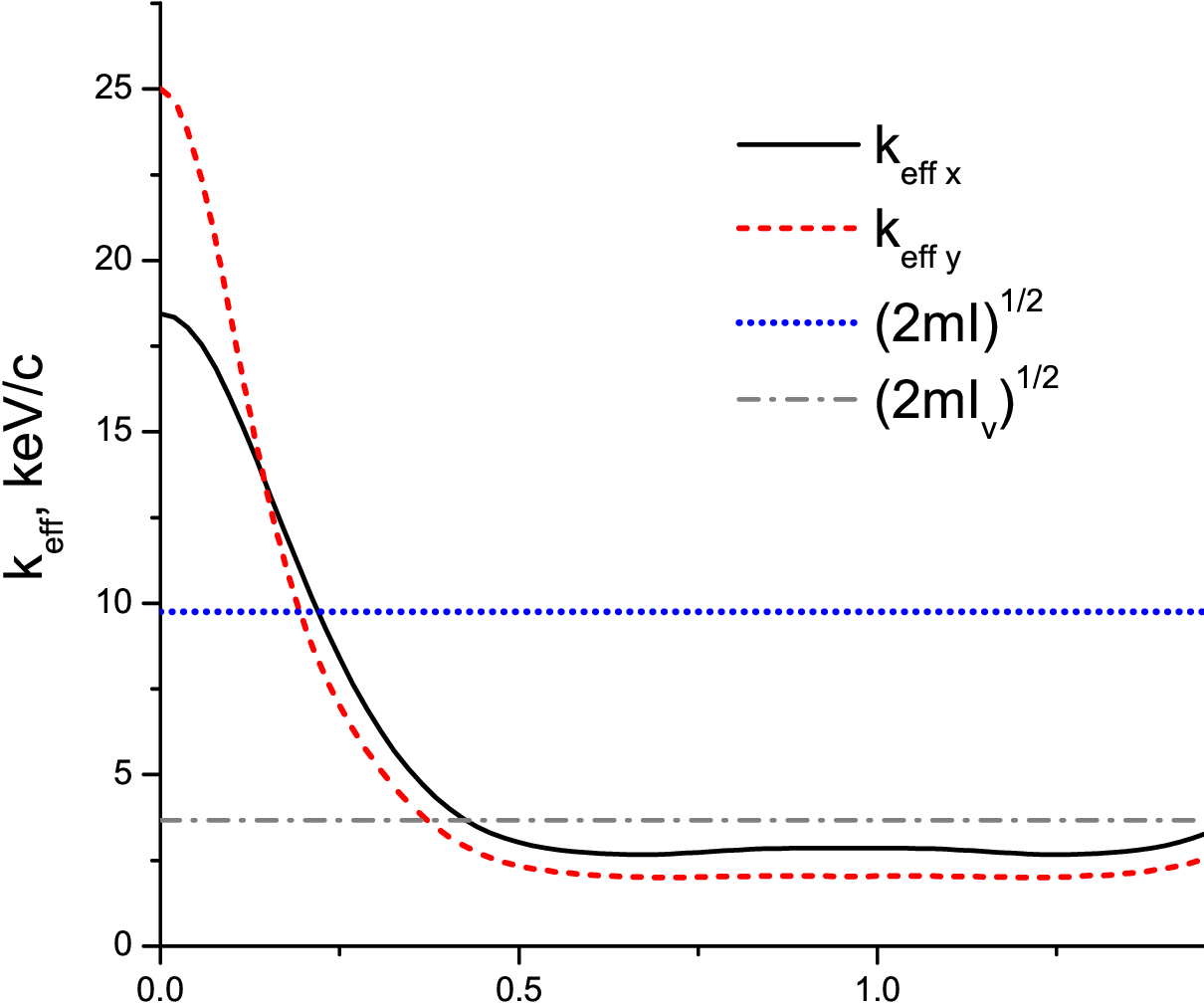}} \caption{The
dependence of the effective minimum momentum on the distance x
from the atomic plane, calculated for the (110) crystallographic
plane of a silicon crystal. The characteristic ionization momentum
and ionization momentum of collectivised electrons are also shown.
}
\end{center}
\end{figure}
Fig. 4 demonstrates that the local effective minimal momenta
$k_{eff,x}(x)$, $k_{eff,y}(x)$ in the region of minimum electron
number density reach the values of 2-3 keV/c, being smaller than
the characteristic ionization momentum (39) and even the
ionization momentum of collectivised electrons (18). This
prediction, which is the main numerical result of this work, is
explained by the fact that the leading role in incoherent
scattering in the region of minimum electron density is played not
by the electrons situated right there, but by electrons located at
distances of the order of the interatomic spacing, the average
concentration of which is nothing other than the average electron
number density in the crystal. At such distances, momentum is
mainly transferred to collectivized electrons with the lowest
excitation energy (18). In addition to this, the effective minimum
transmitted momenta at the center of the interplanar channel are
further reduced due to the algorithm for their introduction (63),
(64) in which the logarithms of the ratios $k_1/k_{eff x}(x)$ and
$k_1/k_{eff y}(x)$ are multiplied by the lowest local electron
number density. Note that even the use of typical scale of local
effective minimal momenta $k_{eff,x}(x)$, $k_{eff,y}(x)$ in the
region of minimum electron density allows one to increase the
accuracy of the description of incoherent scattering of most
stably channeled positively charged particles.

As for the accuracy of calculating the mean square scattering
angle and effective minimum momentum using equations (63) and
(64), it should be taken into account that they are based on a
simplified description of both the substance permittivity and
excitation of collectivized electrons, the characteristics of
which (see the table on page 8) and, in particular, the parameter
$\alpha$ in formula (17), are not precisely determined. Thus, when
going from the value $\alpha = 1/3$, used both in  \cite{lun,
penn, penn2, shin} and in our calculations, to $\alpha = 0.6$ from
\cite{ege, br, fer, rh}, the effective minimum momenta
$k_{eff,x}(x)$ and $k_{eff,y}(x)$ change by about 10\%. However,
the accuracy of the ansatz (58) cannot be reliably assessed at
all. Estimating the total accuracy of expressions (63) and (64),
it should also be taken into account that the maximum momentum of
channeled particles (50) exceeds the effective minimum momentum by
2-3 orders of magnitude. Note also that the predictions of
expressions (63) and (64) can be corrected on the basis of
experimental data by fitting only a single parameter of the
effective ionization energy $I$. It should be noted that fitting
the model of electron scattering in an amorphous substance
(Section III) to experimental data is complicated by the
inevitable presence of nuclear scattering, which is approximately
$Z$ times more intense, increasing thus the importance of studying
electron scattering by channeled particles under conditions of
suppressed nuclear scattering. Presented theory can also be
applied to the scattering of electrons and other negatively
charged channeled particles, however, this will not lead to a
significant refinement of the overall picture due to the same more
intense nuclear scattering that is inevitable in the case of
negatively charged channeled particles.

Both the mean square angles and the effective minimum momenta of
incoherent electron scattering determined by expressions (63) and
(64) can be used as the basis for various methods of probabilistic
modeling of incoheremt electron scatering of channeling particles.

\section{Algorithms for modeling incoherent scattering of high-energy particles on
electrons}

Two Monte Carlo scattering simulation algorithms can be readilly
suggested based on expression (63) and (64). In the first, the
latter are used to calculate the mean square projected scattering
angles in the xz and yz planes for momentum transmissions
$k_\perp< k_1$, while for $k_\perp > k_1$ the scattering is
treated as a single one. For the latter, to provide the most
realistic description of processes in the near-barrier transverse
energy region, the condition
\begin {equation} 
k_1 < \frac{1}{3} p_{ch} = \frac{1}{3}\sqrt{2 V_0 \varepsilon}
\end{equation}
should be imposeed. Furthermore, in order for the local electron
density approximation to be used for modeling the single
scattering, another condition
\begin {equation} 
k_1 > \frac{1}{\delta x}
\end{equation}
must be satisfied, where $\delta x$ is the spatial scale of
variation in the average electron concentration. In a region of
predominant electron scattering, sufficiently distant from the
region of localization of atomic nuclei, we can set $\delta x \sim
5 \times 10^{-10}$ cm and see that both conditions are satisfied
starting from particle energies of several hundred MeV.

Another method can be  based on the sampling of all the electron
scattering events using the screened Coulomb cross section
\begin {equation} 
d^2 \sigma = \frac{4 e^4 d^2k}{v^2(k^2 + \kappa^2_{scr})^2}~,
\end{equation}
which screening parameter $\kappa_{scr}$ is determined by the
equality of mean squared scattering angle calculated by using both
formulas (63), (64) and cross section (72). This condition can be
written in the form
\begin {equation} 
ln \zeta + \frac{1}{\zeta} = a,
\end{equation}
where
\begin {equation} 
\zeta = \frac{k^2_1 + \kappa^2_{scr}}{\kappa^2}
\end{equation}
and
\begin {equation} 
a = 1 + ln \frac{k_1}{k_{eff x,y}(x)}.
\end{equation}
The boundary momentum $k_1$ should be naturally chosen to exseed
the largest momentum important for the considereed problem and
satisfy the condition
\begin {equation} 
k_1 > 3 p_{ch} = 3 \sqrt{2 V_0 \varepsilon},
\end{equation}
insuring the application of the approximate solution
\begin {equation} 
\zeta \simeq e^a -1 - \frac{1}{2}e^{-a} - \frac{2}{3}e^{-2a}
\end{equation}
of equation (73), the accuracy of which is about $10^{-3}$ for $a
= 3$ and rapidly increases with $a$ value. Having found the
screening parameters corresponding to the minimum momenta $k_{eff
x}$ and $k_{eff y}$, one should apply the cross section (72) for
independent sampling of single scattering in the xz and yz planes,
implementing with its help both the local Coulomb scattering
nature at large transferred momenta and the considerable
modification of the Coulomb scattering process at small ones.

\section{Conclusion}
The article presents formulas that make it possible to suggest
algorithms for probabilistic modeling of processes of incoherent
scattering of channeled high-energy particles on electrons of
crystal atoms. First, the modification of the particle
electromagnetic field that appears in a condensed medium at
ultrarelativistic energies is taken into account. When describing
processes not accompanied by coherent transfer of momentum to the
crystal lattice, the given formulas take into account the
excitation of both collectivised electrons and electrons of inner
atomic shells, and when describing processes accompanied by
coherent transmission - the influence of thermal vibrations of
atoms and correlations in scattering by atomic electrons.

Taking these processes into account expressions are obtained for
the mean square projected scattering angles per a unit length
which also make it possible to introduce effective minimum
scattering momenta in planes parallel and perpendicular to the
atomic planes. Expressions of either type make it possible to take
into account the nonlocal effects of the momentum transmission
from crystal electrons to particles moving along classical
trajectories, and to formulate methods for modeling them. In this
case, sampling the process of diffuse scattering of channeled
particles one should take into account only the transmission of
momentum, the transverse component of which does not exceed the
characteristic transverse momentum of channeled particles, while
the processes of transmission of transverse momentum exceeding the
latter should be sampled using both Coulomb cross section and
local electron number density along the particle trajectory.

It is expected that these methods will be used in conjunction with
the advanced modeling of incoherent scattering by nuclei,
described in \cite{tik2} and in \cite{tik3} for the axial and
planar case respectively, also taking into account both the
non-locality of Coulomb scattering and the need to separately
consider diffuse and single processes.

\end{document}